\documentclass[prl,twocolumn,preprintnumbers,amsmath,amssymb]{revtex4-2}
\usepackage{graphicx}
\usepackage{dcolumn}
\usepackage{amsmath,amsbsy,amssymb,scalefnt}
\usepackage{bm}
\usepackage{epstopdf}
\usepackage[T1]{fontenc}
\usepackage{amsfonts}
\usepackage{booktabs}
\usepackage{colortbl}
\usepackage{textcomp}
\usepackage{mathtools}
\usepackage{xcolor}
\usepackage{hyperref}
\usepackage{physics}
\hypersetup{ pdfnewwindow=true, colorlinks=true, linkcolor=blue, anchorcolor=blue, citecolor=blue, filecolor=blue, menucolor=blue, urlcolor=blue}

\begin{document}
\title{Planar Hall effect in ultrathin topological insulator films}

\author{Mohammad Shafiei}
\affiliation{Department of Physics, University of Antwerp, Groenenborgerlaan 171, B-2020 Antwerp, Belgium}

\author{Milorad V. Milo\v{s}evi\'c}
\email{milorad.milosevic@uantwerpen.be}
\affiliation{Department of Physics, University of Antwerp, Groenenborgerlaan 171, B-2020 Antwerp, Belgium}

\date{\today}

\begin{abstract}  
The planar Hall effect (PHE), previously observed in Weyl and Dirac semimetals due to the chiral anomaly, emerges with a different origin in topological insulators (TIs), where in-plane magnetic fields induce resistivity anisotropy. In strictly two-dimensional TIs, PHE is generally suppressed due to the inability of the out-of-plane Berry curvature to couple to the in-plane band velocity of the charge carriers. Here, we demonstrate that in ultrathin TI films, a quasi-two-dimensional system, intersurface tunneling coupling with in-plane magnetization induces electronic anisotropy, enabling a finite PHE. In addition, we reveal that strong in-plane magnetization can stabilize the thickness-dependent quantum anomalous Hall effect, typically associated with out-of-plane magnetization. These insights advance the understanding of magnetic topological phases, paving the way for next-generation spintronic devices and magnetic sensing technologies.
\end{abstract}
\maketitle

\paragraph{Introduction} The planar Hall effect (PHE) is a magnetoresistive phenomenon where an in-plane transverse voltage emerges in response to coplanar electric and magnetic fields~\cite{ghorai2025planar,zhong2023recent,ma2024observation}. 
PHE has garnered significant interest for its potential in high-sensitivity magnetic sensors~\cite{bhardwaj2021observation}, magnetic random access memory devices~\cite{binh2019simple,bason2006planar}, and lab-on-a-chip platforms for detecting low magnetic moments~\cite{elzwawy2021current}. PHE has been realized across various material platforms~\cite{chang2023colloquium,zhong2023recent} including topological metals~\cite{burkov2017giant} and ferromagnetic semiconductors~\cite{zhong2023recent}, with more recent observations in Dirac and Weyl semimetals~\cite{kumar2018planar}, where the chiral anomaly plays a pivotal role. In topological insulators (TIs), PHE emerges through Dirac cone tilting and system-induced anisotropy~\cite{taskin2017planar,zheng2020origin}. In three-dimensional TIs, PHE can serve as a sensitive probe of Berry curvature and topological properties~\cite{ghorai2025planar}. However, in strictly two-dimensional systems, conventional Berry curvature-driven PHE is prohibited due to the inability of out-of-plane Berry curvature to couple to electron velocities confined to the plane~\cite{ghorai2025planar}.

Recent studies have sought to overcome the latter constraint. Specifically, Ghorai et al.~\cite{ghorai2025planar} demonstrated PHE in bilayer graphene, a quasi-two-dimensional system, attributing the effect to interlayer tunneling driven by finite interlayer hopping. Inspired by this study, here we investigate the emergence of PHE in ultrathin TI films as a quasi-two-dimensional platform.  In this regime, strong intersurface coupling induces a hybridization gap, consistent with experimental observations~\cite{zhang2010crossover}. This hybridization reshapes the electronic structure, giving rise to interesting phenomena absent in thicker samples~\cite{shafiei2022controlling,shafiei2024tuning}. Similarly to bilayer graphene, we attribute the emergence of PHE to intersurface tunneling between hybridized TI layers. As intersurface coupling strength can be tuned by varying the film thickness or applying strain~\cite{shafiei2022controlling}, ultrathin TI films present an ideal platform for controlled PHE generation. Furthermore, while Hall responses are typically suppressed in systems lacking strong spin-orbit coupling~\cite{cullen2021generating,liang2018anomalous,ghorai2025planar}, ultrathin TIs exhibit robust spin-orbit interactions, making them a promising candidate for magnetic sensors and advanced two-dimensional devices.

We here demonstrate that intersurface hybridization, coupled with in-plane magnetization, induces electronic anisotropy, enabling a detectable PHE. Our analysis reveals that this coupling arises from second-order momentum contributions within the hybridization gap, emphasizing the nontrivial role of in-plane magnetization. By examining the dependence of PHE on magnetization strength, orientation, and film thickness, we chart out the interplay of magnetic and electronic structure in ultrathin TI films, useful beyond this work alone. 

Additionally, we show that with further increasing in-plane magnetization in this system, a quantum anomalous Hall effect (QAHE) is found, extending its realm beyond the conventional case of out-of-plane magnetization. QAHE characterized by quantized Hall conductance without an external magnetic field~\cite{chang2013experimental}, enables dissipationless edge current transport — a key ingredient for low-power electronics, quantum metrology, and topological quantum computation~\cite{tokura2019magnetic,chang2023colloquium}. QAHE was first observed in magnetically doped TIs~\cite{chang2013experimental} and later achieved in TI/ferromagnet heterostructures~\cite{xu2015quantum}. Our results delve deeper into the understanding of magnetic topological insulators in the ultrathin limit, and add to their potential for next-generation spintronics, quantum computing technologies, and Hall sensors, where precise control of electronic and topological properties is paramount.

\paragraph{System and Hamiltonian} As a generic example, we consider an ultrathin Bi$_2$Se$_3$ film doped with magnetic atoms or subjected to an external magnetic field. This family of materials is well established for their topological properties and rhombohedral crystal structure~\cite{zhang2009topological}, with a unit cell comprising two Bi atoms and three Se atoms known as a quintuple layer (QL)~\cite{zhang2009topological,zhang2010crossover}. The effective low-energy Hamiltonian for the surface states of this system can be described as~\cite{zhang2010crossover,liu2010model,zheng2020origin}:
\begin{equation}
H_{TI} = \sum_{k'\sigma\sigma'} c^\dagger_{k'\sigma} h_{TI}(k')_{\sigma\sigma'} c_{k'\sigma'},
\end{equation}
where
\begin{equation}\label{hamiltonian}
h_{TI}(k') = (\Delta_0-\Delta_1k'^2) \sigma_z + v_F (k' \times \sigma)_z - \mathbf{\sigma} \cdot \mathbf{M}. 
\end{equation}
Here, $c^\dagger_{k'\sigma}$ ($c_{k'\sigma'}$) are the creation (annihilation) operators of electrons, $v_F$ is the Fermi velocity, $\mathbf{\sigma}$ are Pauli matrices for spin, $k'=(k'_x,k'_y)$ is the in-plane momentum, and $\mathbf{M}$ is the in-plane magnetization vector, induced by magnetic adatoms or an external in-plane magnetic field. For magnetically doped TIs, magnetic impurities distributed over the lattice sites produce a net in-plane magnetization $M$, with higher concentrations leading to increased magnetization. For an applied in-plane magnetic field, the resulting magnetization arises from the Zeeman exchange interaction induced by the field. Thus, the effect of magnetic impurities or the magnetic field is included as the Zeeman exchange term in the Hamiltonian (Eq.~\eqref{hamiltonian}). The parameters $\Delta_0$, $\Delta_1$, and $v_F$ depend strongly on film thickness. We compiled experimental data for these parameters at various thicknesses~\cite{zhang2010crossover}, summarized in Table~\ref{table1}. Both $\Delta_0$ and $\Delta_1$ decrease with increasing thickness and vanish beyond the 5QL thickness, marking the transition from the ultrathin to bulk regime~\cite{zhang2010crossover,liu2010oscillatory}.

\begin{table}[b]
\centering
\begin{tabular}{|c|c|c|c|c|c|}
\bottomrule
\bottomrule
\textbf{\,\,\,\,\,\,\,\,\,\,\,\,\,\,\,\,\,\,\,\,} & \textbf{2QL} & \textbf{3QL} & \textbf{4QL} & \textbf{5QL} & \textbf{6QL} \\
\midrule
$\Delta_0$ (eV) & 0.126 & 0.069 & 0.035 & 0.020 & 0 \\
$\Delta_1$ (eV $\textup{\AA}^2)$ & 21.8 & 18 & 10 & 5 & 0 \\
$v_F$ (eV $\textup{\AA}$) & 3.10 & 3.17 & 2.95 & 2.99 & 2.98\\
\bottomrule
\bottomrule
\end{tabular}
\caption{Thickness-dependent parameters $\Delta_0$, $\Delta_1$, and $v_F$ for ultrathin TI films (2-6QL). Values derived from experimental data~\cite{zhang2010crossover}. 
}
\label{table1}
\end{table}

In-plane magnetization $\mathbf{M}$ is given by $\mathbf{M} = (M_x, M_y) = (M\cos\theta, M\sin\theta)$, where $\theta$ is the angle between magnetization and the $x$-axis. For samples thicker than 5QL, the parameter $\Delta_1$ vanishes, and the presence of $\mathbf{M}$ only induces a shift in the Dirac cone. However, for samples with thicknesses of 5QL or less, the second-order momentum term from the hybridization gap shifts the Dirac cone's origin $(0,0) \to (M_y/v_F, -M_x/v_F)$ and introduces anisotropy in its dispersion. This effect is evident when the Hamiltonian is expanded around the shifted Dirac point at $(M_y/v_F, -M_x/v_F)$, as follows:

\begin{equation}\label{total_hamiltonian}
    \begin{aligned}
    H_{\text{TI}}(\mathbf{k})& = \left(\Delta_0 - \Delta_1 k^2 - \frac{\Delta_1 M^2}{v_F^2} \right) \sigma_z  
+ v_F (\mathbf{k} \times \boldsymbol{\sigma})_z\\  
        &- \frac{2M \Delta_1}{v_F} (k_x \sin \theta - k_y \cos \theta) \sigma_z.
    \end{aligned}
\end{equation}
As evident from Eq.~\eqref{total_hamiltonian}, the coupling between in-plane magnetization and the hybridization gap not only introduces a mass term of magnitude $\Delta_1 M^2/v_F^2$ to the Dirac cone but also induces anisotropy in momentum space, characterized by $2M \Delta_1 / v_F (k_x \sin\theta-k_y\cos\theta)$.

\begin{figure}[b]
    \centering
     \includegraphics[width=0.8\linewidth]{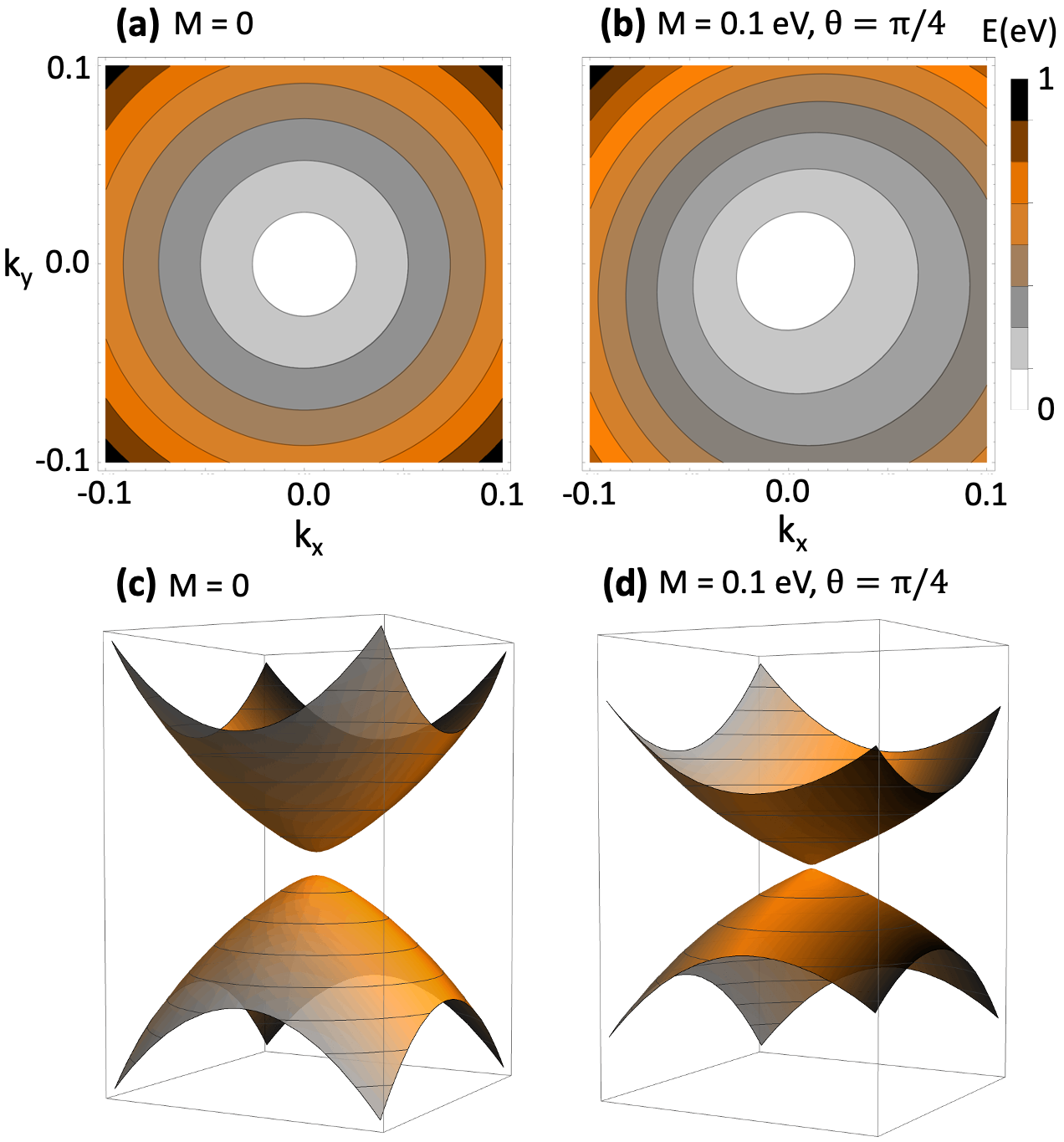}
    \caption{Contour plots of the conduction band quasienergy $E$ in momentum space for (a) the nonmagnetic system and (b) the system with in-plane magnetization, for 3QL thickness of the film. Dirac energy band structure of an ultrathin TI (c) in the absence and (d) in presence of in-plane magnetization. The comparison shows that in-plane magnetization not only modifies the hybridization gap but also induces a significant deformation in both the Fermi energy contours and the Dirac energy band structure.}
    \label{fig:fig1}
\end{figure}

\begin{figure*}[t]
    \centering
     \includegraphics[width=0.95\linewidth]{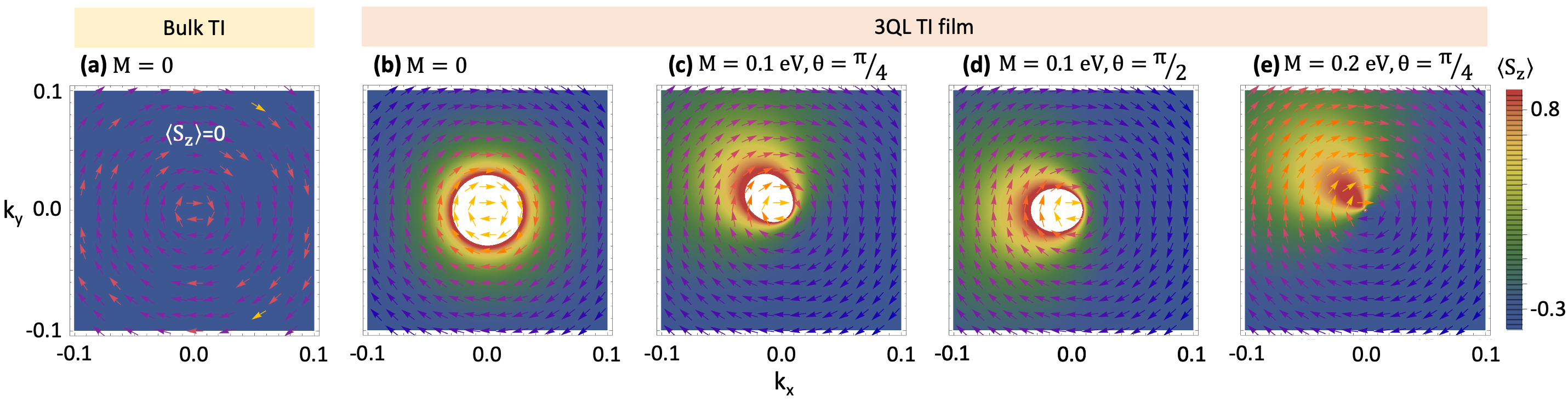}
    \caption{Spin texture in momentum space for surface states of a TI in the conduction band, shown for thick ($\Delta = 0$) and ultrathin ($\Delta \neq 0$) regimes. (a) In the absence of a hybridization band gap, the spin is locked to the momentum with only in-plane components. (b) In the ultrathin regime with nonzero band gap (here for thickness 3QL, cf. Table~\ref{table1}), both in-plane and $z$-component spins appear. (c-e) The effect of in-plane magnetization magnitude ($M = 0.1,0.2$ eV) and direction ($\theta = \pi/4, \pi/2$) on the spin texture, demonstrating tunability.}
    \label{fig:fig2}
\end{figure*}

\paragraph{Band structure and spin texture} We first analyze the band structure and spin texture of an ultrathin TI with in-plane magnetization, highlighting how the coupling between the hybridization gap and in-plane magnetization in the ultrathin regime affects these properties. For the effective Hamiltonian described by Eq.~\eqref{total_hamiltonian}, the energy eigenvalues are given by:
\begin{equation}
    \begin{aligned}
    E_{k\gamma} &= \gamma \sqrt{\left(\Delta_0 - \Delta_1 k^2 - D  \right)^2 + v_F^2 k^2},\\
    D & =  \frac{\Delta_1 M^2}{v_F^2} + \frac{2M \Delta_1}{v_F} (k_x \sin \theta - k_y \cos \theta),
    \end{aligned}
\end{equation}
where $\gamma=\pm$ corresponds to the conduction and valence bands. Figure~\ref{fig:fig1} shows the Dirac energy band structure for both the nonmagnetic and in-plane magnetized systems with 3QL thickness. The contour plots of the quasi-energy of the conduction band $E_+$ are also displayed in the momentum space. A comparison between the left and right panels of Fig.~\ref{fig:fig1} reveals that in-plane magnetization modifies the hybridization gap at the $\Gamma$ point and significantly deforms the Fermi energy contours and Dirac energy band. 

We also investigated the spin texture of the surface states for an in-plane magnetized TI. The spin texture is the vector field in momentum space representing the expectation values of the spin orientations, calculated as:
\begin{equation}
\vec{S} = \frac{1}{2} \langle u_k | \vec{\sigma} | u_k \rangle = \frac{1}{2}[\langle\sigma_x\rangle \hat{x} + \langle\sigma_y\rangle \hat{y} + \langle\sigma_z\rangle \hat{z}],
\end{equation}
where $u_k$ is the eigenstate of the effective Hamiltonian in Eq.~\eqref{total_hamiltonian}. Figure~\ref{fig:fig2} shows the spin texture in momentum space for the surface states in both thick ($\Delta_{0,1} = 0$) and ultrathin ($\Delta_{0,1} \neq 0$) regimes, specifically within the conduction band. In the absence of a hybridization gap, Fig.~\ref{fig:fig2}(a) shows that, at low energies, only in-plane spin components are present, and the spin is locked to the momentum. This is linked to the surface state Hamiltonian, which, without $\Delta_0$ and $\Delta_1$, involves only the $\sigma_x$ and $\sigma_y$ Pauli matrices. In the ultrathin film regime, where $\Delta_0 \neq 0$ and $\Delta_1 \neq 0$, the Hamiltonian includes the $\sigma_z$ term, leading to a spin texture (Fig.~\ref{fig:fig2}(b)) that exhibits both in-plane and $z$-component spins.  When in-plane magnetization is introduced, the spin texture is no longer isotropic, as shown in Fig.~\ref{fig:fig2}(c-e). The effect is demonstrated for magnetization values of $M = 0.1, 0.2$ eV at angles $\theta = \pi/4$ and $\pi/2$. The in-plane magnetization thus enables tuning of the spin texture by adjusting either the magnitude ($M$) or direction (angle $\theta$) of the magnetization.

\begin{figure}[b]
    \centering
     \includegraphics[width=0.98\linewidth]{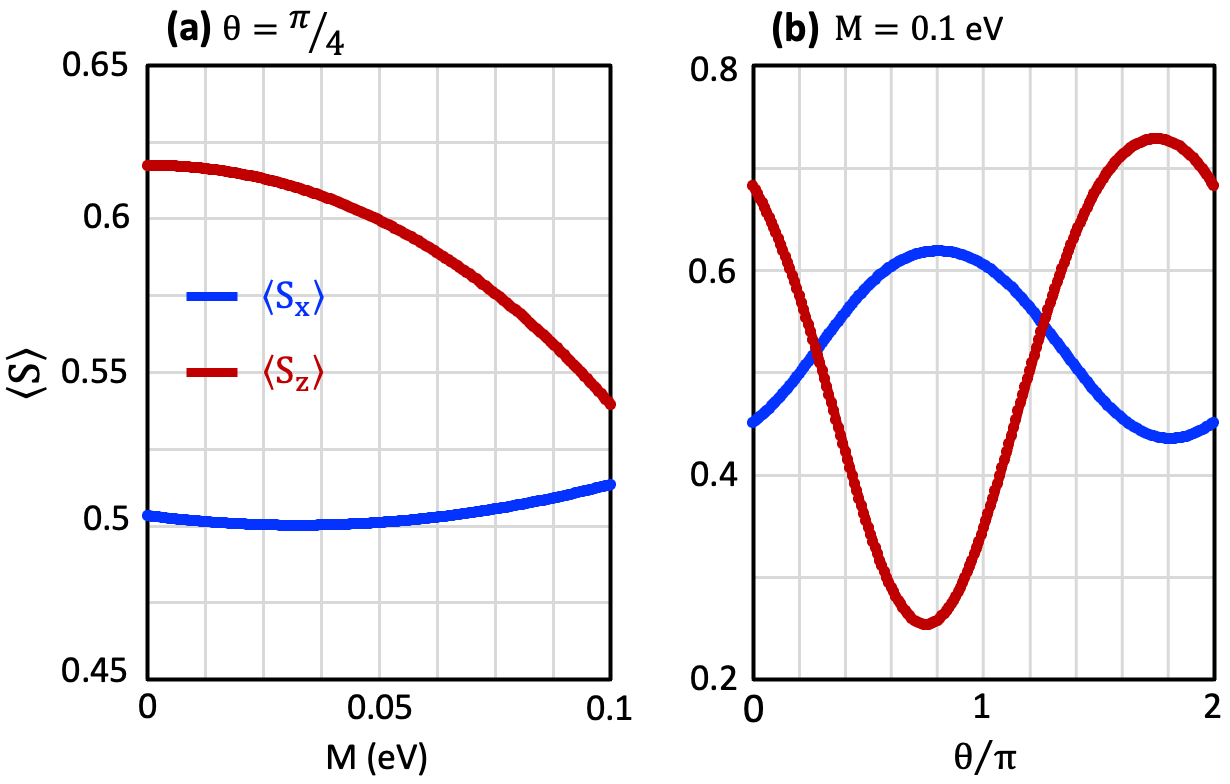}
    \caption{Average spin components $\langle S_x \rangle$ and $\langle S_z \rangle$ as a function of (a) magnetization magnitude $M$ and (b) magnetization angle $\theta$, for a 3QL TI film.}
    \label{fig:fig3}
\end{figure}

\begin{figure*}[t]
    \centering
     \includegraphics[width=0.75\linewidth]{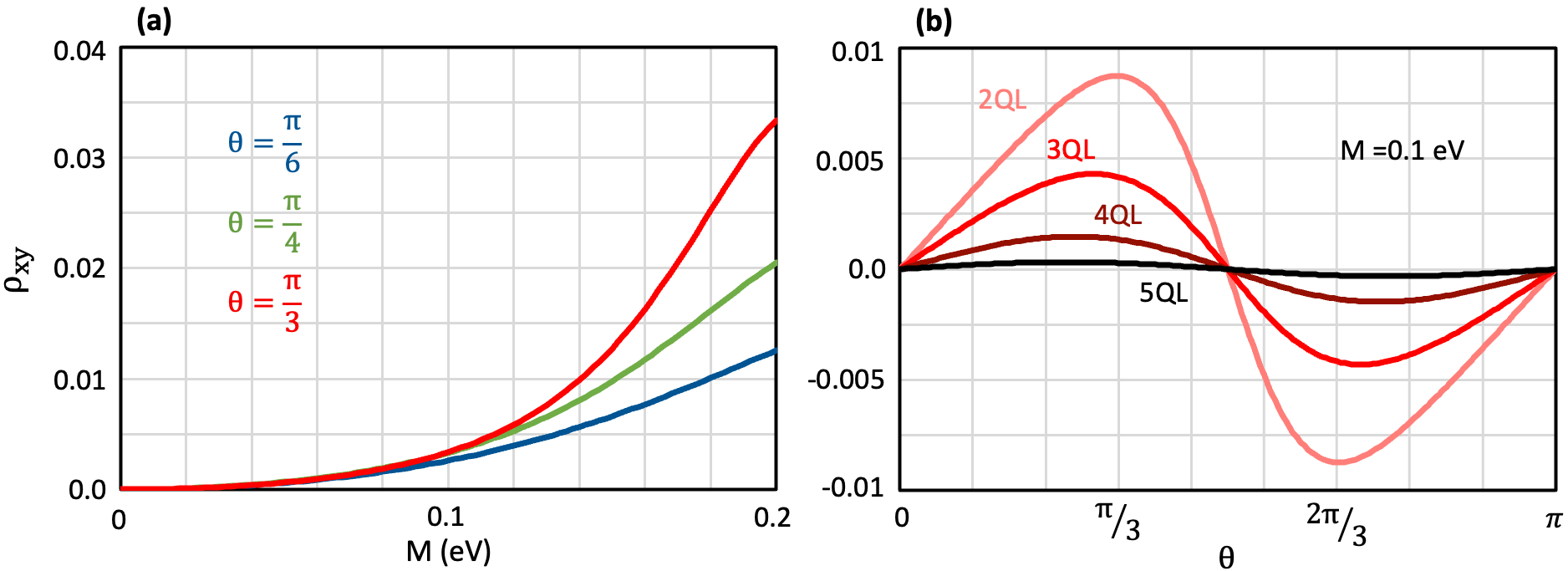}
    \caption{Calculated Hall resistivity ($\rho_{xy}$) as a function of (a) in-plane magnetization magnitude (3QL film) and (b) orientation angle $\theta$ (2QL-5QL films). The plot demonstrates a direct correlation between magnetization magnitude and $\rho_{xy}$, while the oscillatory behavior with respect to $\theta$ indicates the influence of magnetic anisotropy on the Hall resistivity.}
    \label{fig:fig4}
\end{figure*}

Moreover, in Fig.~\ref{fig:fig3}, we calculate the average spin components $\langle S \rangle$ along the $x$ and $z$ directions as functions of in-plane magnetization $M$ and the angle $\theta$. As shown in panel (a), for constant $\theta$, increasing $M$ reduces $\langle S_z \rangle$, which can be attributed to the decrease in the hybridization gap, caused by the term $-\Delta_1 M^2/v_F^2$. Simultaneously, $\langle S_x \rangle$ slightly increases. Panel (b) shows that when the magnetization magnitude is fixed and $\theta$ is varied, $\langle S_z \rangle$ and $\langle S_x \rangle$ oscillate due to the anisotropy introduced by the last term in the Hamiltonian (Eq.~\eqref{total_hamiltonian}).

\paragraph{PHE and QAHE in ultrathin TIs} Having understood the background, we next examine the PHE and QAHE in ultrathin TI films with in-plane magnetization. The PHE originates from an anisotropic term in the Hamiltonian, which leads to a non-topological Hall response. This effect results from the interaction of the system's electronic states with an in-plane magnetic field, leading to longitudinal and transverse voltage generation. In contrast, the QAHE arises from the intrinsic topology of the electronic band structure and manifests through the formation of chiral edge states. First we investigated the realization of  PHE arising from the resistivity anisotropy induced by in-plane magnetization in an ultrathin TI.
We adopt the Streda-Smrcka formalism~\cite{streda1982theory,sinitsyn2006charge,zheng2020origin} within the Kubo formula to calculate the conductivity:
\begin{equation}
    \sigma_{ij} = - \frac{e^2}{2h} \int d\omega \frac{\partial f(\omega)}{\partial \omega} \sum_{\mathbf{k}} \text{Tr} \big\{ v_i [ G^r (\mathbf{k}, \omega) - G^a (\mathbf{k}, \omega) ] \big.
\end{equation}
\begin{equation*}
    \big. \times v_j G^a (\mathbf{k}, \omega) - v_i G^r (\mathbf{k}, \omega) v_j [ G^r (\mathbf{k}, \omega) - G^a (\mathbf{k}, \omega) ] \big\},
\end{equation*}
where $v_i = \partial h_{\text{TI}} / \partial k_i$ represents the velocity operator, while $f(x) = [e^{(x-\mu)/k_B T} + 1]^{-1}$ denotes the Fermi distribution function. The retarded and advanced Green’s functions, $G^{r/a} (\mathbf{k}, \omega)$, take the form:
\begin{equation}
    G^{r/a} (\mathbf{k}, \omega) = \frac{1}{2} \sum_{\gamma = \pm} [1 + \gamma \mathbf{n_k} \cdot \bm{\sigma}] G^{r/a}_{\gamma} (\mathbf{k}, \omega).
\end{equation}
In this formulation, the momentum unit vector is given by $\mathbf{n_k} = (-k_y, k_x, 0)/|\mathbf{k}|$. 

To elucidate the impact of the modified band structure, a perturbative approach is adopted, initiating with the introduction of a weak, constant self-energy term, $i\Sigma$, at zero temperature, thereby simulating a weak scalar potential. This method is comprehensively described in Ref.~\cite{zheng2020origin}. Notably, the anomalous Hall conductivity, $\sigma_{xy}$, exhibits symmetry with respect to the magnetization, indicating a mechanism distinct from the conventional Lorentz force. 
Figure~\ref{fig:fig4} presents the computed Hall resistivity, $\rho_{xy} = \sigma_{xy} / (\sigma_{xx}^2 + \sigma_{xy}^2)$, as a function of in-plane magnetization magnitude and orientation angle $\theta$. The results demonstrate a clear trend: an increase in magnetization magnitude correlates with an increase in $\rho_{xy}$. Furthermore, the Hall resistivity exhibits oscillatory behavior as a function of $\theta$, reflecting the influence of anisotropy. 
The observed Hall effect can be rationalized by considering the role of spin-flip scattering. In the absence of band modifying, the energy dispersion, $E_{k\gamma}$, remains isotropic with respect to the momentum angle, resulting in the cancellation of the spin-flip contribution, $\mathbf{n}_{k\gamma} \cdot \mathbf{\sigma}$, upon angular integration. However, the introduction of a finite change induces an angular dependence in $E_{k\gamma}$ proportional to $M_x k_y - M_y k_x$, disrupting this cancellation. Physically, the inherent prohibition of backscattering between states $|k\rangle$ and $|-k\rangle$ is lifted along the field direction, a consequence of the non-orthogonal spin orientations, $s_k$ and $s_{-k}$, arising from the elliptical deformation of the Fermi surface, as depicted in Fig.~\ref{fig:fig1}.

We also investigated the QAHE arising from the coupling between intersurface tunneling and in-plane magnetization in ultrathin TIs. In thick TIs, out-of-plane magnetization or a mass term for the Dirac cone breaks time-reversal symmetry, leading to QAH or Chern insulator states~\cite{chang2023colloquium}, characterized by the topological invariant $C$, representing the number of chiral edge states. In ultrathin magnetic TIs, two energy gaps emerge: the hybridization gap, from coupling of top and bottom surface states, and the magnetic exchange gap induced by magnetization~\cite{shafiei2022controlling,shafiei2024tuning}. The competition between these gaps determines the system's topological phase. When the magnetic exchange gap dominates, the QAH state with Chern number $C = \pm 1$ appears, while a dominant hybridization gap results in a trivial state with $C = 0$. Thus, a strong magnetic exchange gap is required for the QAH state.

\begin{figure}[t]
    \centering
     \includegraphics[width=0.85\linewidth]{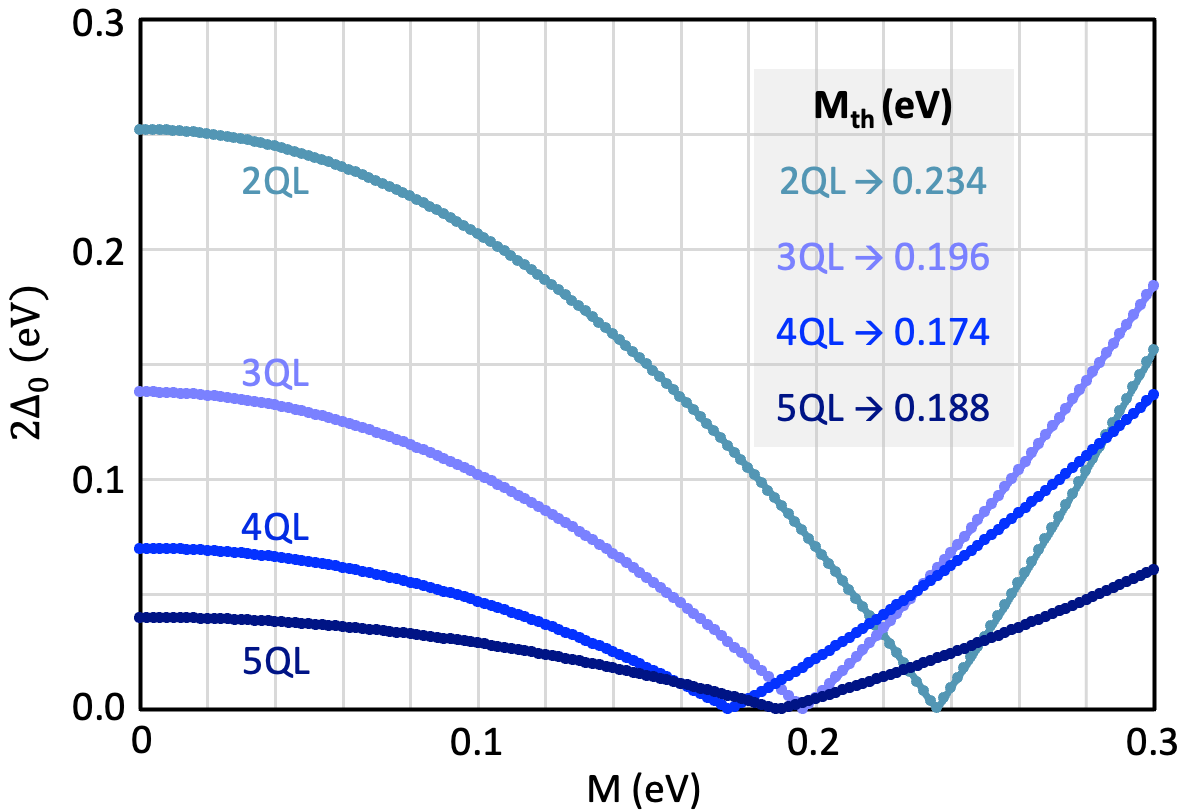}
    \caption{Dependence of the hybridization gap on in-plane magnetization for ultrathin TIs with thicknesses of 2 to 5 QL. For magnetization values above $M_{\text{th}}$, the system transitions into the QAH phase with a Chern number $C = -1$, leading to a quantized Hall conductance of $-e^2/h$.}
    \label{fig:fig5}
\end{figure}

In ultrathin TIs with in-plane magnetization, the coupling between the magnetization and surface states induces a mass term $-\Delta_1 M^2 / v_F^2$ in Eq.~\eqref{total_hamiltonian}, acting as a magnetic exchange gap. This reduces the hybridization gap, and when large enough, can induce a topological phase transition into the QAH state by closing and reopening the gap. Figure~\ref{fig:fig5} shows how the hybridization gap varies with in-plane magnetization. The threshold magnetization ($M_{\text{th}}$) for the QAH state depends on thickness, with values of 0.234, 0.196, 0.174, and 0.188~eV for samples with 2 to 5 QLs, respectively. For magnetization above these thresholds, the system enters the QAH phase, with chiral edge states. Due to the negative magnetic exchange gap, the system hosts a QAH state with Chern number $C = -1$, and the Hall conductance is quantized as $\sigma_{xy} = -e^2 / h$. In contrast to thicker TIs, a sufficiently strong in-plane magnetization alone can induce the QAHE in ultrathin TIs.

\paragraph{Conclusion} In conclusion, we have demonstrated that ultrathin topological insulator (TI) films, characterized by strong intersurface hybridization, exhibit a planar Hall effect (PHE) when subjected to in-plane magnetization. PHE stems from the magnetization-induced electronic anisotropy in this system, overcoming the conventional restrictions in strictly two-dimensional TIs. Such an anisotropic transport response to coplanar electric and magnetic fields in quasi-two-dimensional topological materials, besides generating PHE, further expands the already rich realm of magnetism-induced phenomena in topological systems~\cite{tokura2019magnetic,bernevig2022progress}.

Furthermore, we demonstrated that a sufficiently strong in-plane magnetization can induce an effective mass in the Dirac spectrum, leading to a stabilized quantum anomalous Hall effect (QAHE) — a behavior traditionally linked to the presence of out-of-plane magnetization. Our findings thus present an alternative mechanism for realizing such magnetic topological phases in ultrathin TIs, beyond the established frameworks.

Having provided deeper insights into the coupling between magnetization, hybridization, and electronic structure in ultrathin TIs, we argue that thin TI films are a particularly convenient tunable platform for exploring (magnetic) topological phases. The here reported planar Hall response in such systems not only contributes to fundamental understanding but also offers practical implications — such as the diagnostic tool for identifying topological phase transitions~\cite{ma2024observation}, distinguishing bulk vs. surface conduction~\cite{nandy2018berry}, and providing precise control over current modulation based on the magnitude and orientation of the magnetic field. These features make PHE valuable for developing spin-based sensors, programmable magnetoresistive logic, and platforms for Majorana mode manipulation in quantum technology~\cite{liu2021magnetic,rao2021theory}.

\paragraph{Acknowledgments} This research was supported by the Research Foundation-Flanders (FWO-Vlaanderen), the FWO-FNRS EOS project ShapeME, and the Special Research Funds (BOF) of the University of Antwerp.

\bibliographystyle{apsrev4-1}
\bibliography{bibliography}

\end{document}